\documentclass{segabs}
\usepackage{multicol}
\usepackage{lipsum}
\usepackage{graphicx}
\usepackage{amssymb,amsmath,amsthm}
\usepackage{amssymb}
\usepackage{amsmath}
\usepackage{caption}
\usepackage{url}
\usepackage{float}
\usepackage{subfigure}
\usepackage{graphicx}  
\usepackage{color}
\newcommand{\VD}[1]{\textcolor{black}{#1}}

\begin{document}

\title{Large-scale frequency-domain seismic wave modeling on {\it{h}}-adaptive tetrahedral meshes with iterative solver and multi-level domain-decomposition preconditioners}
\renewcommand{\thefootnote}{\fnsymbol{footnote}} 

\author{
        V. Dolean \footnotemark[1],
        P. Jolivet\footnotemark[2],
        S.~Operto\footnotemark[3],
        P.H. Tournier\footnotemark[4]  
         \newline
\footnotemark[1] $~$ Univ. of Strathclyde, United Kingdom/UCA, CNRS, LJAD, France; \\
\footnotemark[2] $~$ IRIT-CNRS, University of Toulouse, France; \\
\footnotemark[3] $~$ UCA, CNRS, Geoazur, France;\\
\footnotemark[4] $~$ Sorbonne University, CNRS, France;
        }
\maketitle

\lefthead{Dolean et al.}
\righthead{frequency-domain seismic wave solvers}
\renewcommand{\thefootnote}{\fnsymbol{footnote}} 
%
%
%
\begin{abstract}
Frequency-domain full-waveform inversion (FWI) is suitable for long-offset stationary-recording acquisition, since reliable subsurface models can be reconstructed with a few frequencies and attenuation is easily implemented without computational overhead. In the frequency domain, wave modeling is a Helmholtz-type boundary-value problem which requires to solve a large and sparse system of linear equations per frequency with multiple right-hand sides (sources). This system can be solved with direct or iterative methods. While the former are suitable for FWI application on 3D dense OBC acquisitions covering spatial domains of moderate size, the later should be the approach of choice for sparse node acquisitions covering large domains (more than 50 millions of unknowns). Fast convergence of iterative solvers for Helmholtz problems remains however challenging \VD{in high frequency regime} due to the non definiteness of the Helmholtz operator, \VD{on one side and on the discretization constraints in order to minimize the dispersion error for a given frequency, on the other side}, hence requiring efficient preconditioners. In this study, we use the Krylov subspace GMRES iterative solver combined with a \VD{two-level} domain-decomposition preconditioner. Discretization relies on continuous \VD{Lagrange} finite elements \VD{of order 3} on unstructured tetrahedral meshes to comply with complex geometries and adapt the size of the elements to the local wavelength ($h$-adaptivity). We assess the accuracy, the convergence and the scalability of our method with the acoustic 3D SEG/EAGE Overthrust model up to a frequency of 20~Hz.
\end{abstract}
%
%
\vspace{-0.5cm}
\section{Introduction}
\vspace{-0.3cm}
The ocean bottom node (OBN) acquisition is emerging for deep-offshore seismic imaging by full waveform inversion (FWI) \citep{Beaudoin_2007_FDO}.
These stationary-recording acquisitions have the versatility to design ultra-long offset surveys, which provide a wide angular illumination of the subsurface amenable to broadband velocity models.
This wide-angle illumination allows for efficient frequency-domain (FD) FWI by decimating the multi-fold wavenumber coverage through a coarse frequency sampling  \citep{Pratt_1999_SWIb}. This frequency subsampling makes FD modeling competitive with time-marching methods and leads to compact datasets \citep{Plessix_2017_CAT}. 
Moreover, attenuation is easily implemented in FWI without computational overheads, even improving the \VD{spectral properties (which condition the behavior of solvers)} of Helmholtz operators. In this context, we present a new solver for 3D FD wave simulation as a forward engine for FWI. \VD{Note that a similar solver has been successfully used in solving a medical imaging problem \citep{Tournier:2019:MTI}}. 
FD seismic modeling is a boundary-value problem, \VD{that after discretization by a finite element method, for example, leads to} a sparse linear system whose unknown is the wavefield, the right-hand side (RHS) the seismic source and the coefficients embed the subsurface properties. \VD{To solve such a linear system, one can use either a {\it{sparse direct solver}} \citep{Duff:2017:DMS} or an iterative solver \citep{Saad_2003_IMS}. While a direct solver is efficient when processing multiple RHSs for problems of moderate size ($< 50.10^6$ unknowns) \citep{Amestoy:2016:FFF}, the memory overhead generated by the storage of the LU factors and the limited scalability of the LU decomposition, makes application on large scale problems challenging}. The second approach relies on {\it{iterative solvers}}, whose natural scalability and moderate memory demand make them suitable for large-scale problems. However, {one major issue is}  the convergence speed of iterative solvers when applied to Helmholtz problems \VD{and this convergence deteriorates as the frequency increases. The use of an iterative method, depends on an efficient preconditioner} with the ultimate goal to make the iteration count independent to frequencies, and the processing of multiple RHSs. \\
Here, we focus on the second category because we target large computational domains (several hundred of millions of unknowns) with a limited number of reciprocal sources (from few hundreds to few thousands). Our method relies on a finite-element discretization on a tetrahedral mesh with \VD{Lagrange elements of order 3 whose dispersion properties are improved compared to the 27-points finite difference scheme from \cite{Operto_2007_FDFD,Operto_2014_FAT,Gosselin_2014_FDF}}, the Krylov subspace GMRES solver \citep{Saad_2003_IMS} and a Schwarz two-level domain decomposition preconditioner \citep{Graham:2017:RRD,Bonazzoli:2019:ADD}. Compared to the \VD{well-known} preconditioner based upon shifted Laplacian \citep{Erlangga:2008:AIM}, it is less sensitive to the shift (added attenuation) and can be used without it.
In the following, we briefly review the \VD{discretization and then the solution method}, before assessing the strong and weak scalability of the solver on the 3D SEG/EAGE Overthrust model. 
%
%
\vspace{-0.3cm}
\section{Tetrahedral Finite element method}
\begin{figure}[ht!]
\begin{center}
\includegraphics[width=7.5cm,clip=true]{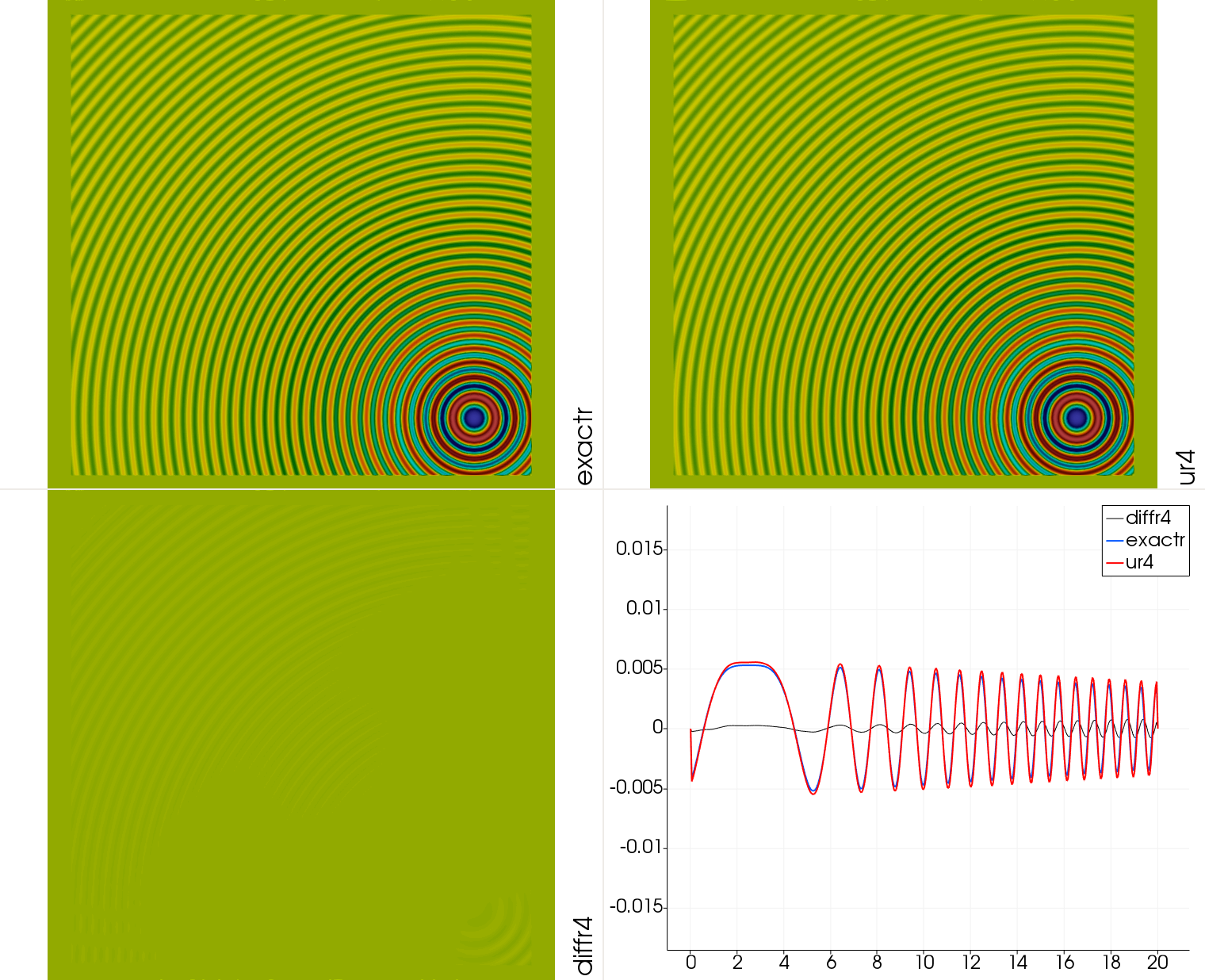}
\caption{Monochromatic wavefield in infinite homogeneous medium. (a) Analytical solution. (b) P2 FE solution with 10 ppwl. (c) Difference. (d) Profile across (a), (b) and (c). Only small amplitude discrepancies are shown.}
\label{compP2}
\end{center}
\end{figure}

\begin{figure}[ht!]
\begin{center}
\includegraphics[width=7.55cm,clip=true]{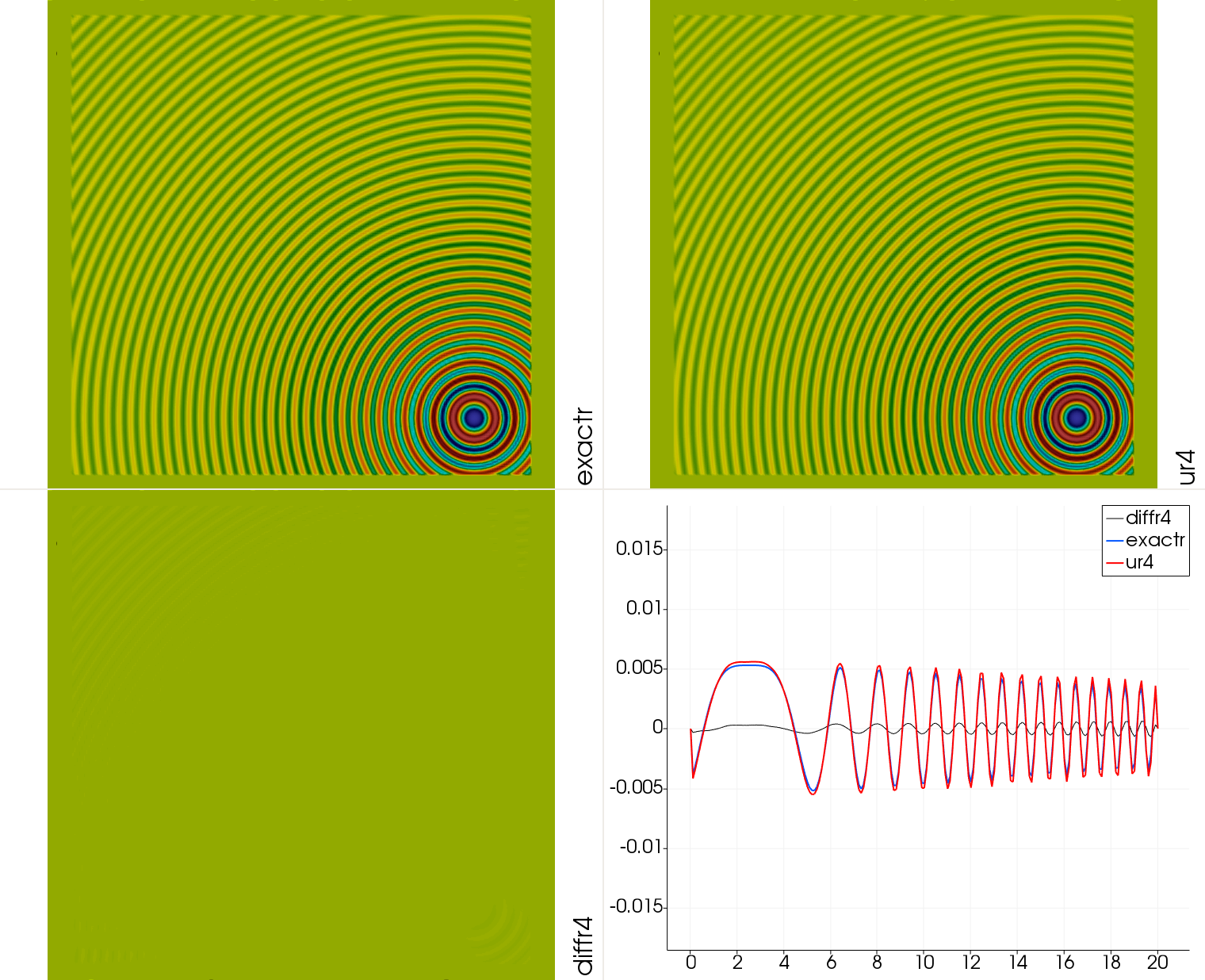}
\caption{Same as Fig. \ref{compP2} for P3 elements and 5 ppwl. An accuracy similar to that achieved by P2 and 10ppwl is shown.}
\label{compP3}
\end{center}
\end{figure}

\vspace{-0.3cm}
The mathematical model of acoustic wave propagation is the Helmholtz equation
\begin{equation}
\left(\Delta + k^2(\bold{x}) \right) u(\bold{x},\omega) = b(\bold{x},\omega), ~ \text{in a subsurface domain } \Omega,
\label{eqh}
\end{equation}
where $u$ is the monochromatic pressure wavefield, $b$ the mono-chromatic source, $k(\bold{x},\omega)=\omega/c(\bold{x})$, with $\omega$ denoting frequency, $c(\bold{x})$ the wavespeed (which is complex valued in viscous media) and $\bold{x}=(x,y,z) \in \Omega$. \\
After discretization, eq. (\ref{eqh}) can be written in matrix form as
\begin{equation}
\bold{A} \bold{u} = \bold{b}.
\label{eq1}
\end{equation}
We implement the above equation with absorbing boundary conditions along the vertical and bottom faces of $\Omega$ and a homogeneous Dirichlet condition on the pressure along the top face (i.e., free-surface boundary condition). \\
We discretize the Helmholtz equation, eq.~\eqref{eqh}, with Lagrange finite elements (FE) on a tetrahedral mesh $\Gamma$ of the domain  $\Omega$.  The rationale behind this choice is multiple: Compared to finite-difference methods on uniform Cartesian grid \citep{Operto_2014_FAT,Gosselin_2014_FDF}, the flexibility offered by unstructured meshes to adapt the size of the elements to the local wavelengths (the so-called $h$-adaptivity) offers a good trade-off between the precision and the number of degrees of freedom (d.o.f) in the mesh. This is particularly true for elastic wave simulation where the shear wavespeeds can reach very low values just below the sea bottom.
Also, compared to hexahedral meshes used with the spectral element method \citep{Li_2020_FEW}, tetrahedral elements are more versatile to conform the mesh to complex known boundaries (topography, sea bottom, salt bodies) and refine the discretization when FWI proceeds over different frequency bands. \\
We conduct a series of numerical experiments to find the degree of Lagrange polynomial providing a sufficient accuracy for a discretization rule of \VD{five} grid points per wavelength (ppwl). 
We solve the  Helmholtz system (\ref{eq1}) with the preconditioned GMRES solver (see next section for more details) for an infinite homogeneous $20km \times 20km \times 4.65km$ medium and a 500-m wavelength and compare the solution with the analytical solution.  Polynomials of degree $3$ (P3) and \VD{$5$} ppwl (Figure \ref{compP3}) achieve the same level of precision than polynomials of degree $2$ (P2)  and $10$ ppwl (Figure \ref{compP2}). Compared to P2, the P3 discretization also implies a reduction from $206$ million to $74$ million d.o.f. and a reduction of the solution time from 93s to 24s on 768 cores. We therefore choose to use the P3 elements in the next section to assess the performances of the parallel preconditioner. The choice of P3 elements is consistent with the conclusions of \citet{Chaumont-Frelet_2016_HMH} for Helmholtz problem and \citet{Mulder_2019_SML} for time-domain simulations on tetrahedral meshes.  \\
These numerical results are also supported by the dispersion analysis of \cite{Ainsworth:2010:OBS}, from which we plot the normalized numerical phase velocity against $1/G$, with $G$ the number of points per wavelength. The dispersion curve shows that the P3 elements provide a better accuracy over a wide range of wavelengths than the 27-point finite-difference stencil of \citet[][, see their Figure 3a]{Operto_2014_FAT}, while the accuracy of the P2 elements is clearly insufficient for \VD{$G=5$}. Note however that the number of non-zero coefficients per row in $\bold{A}$ ranges between 47 and 217 for P3 elements, while this number is 27 for the finite-difference discretization. \\
The simulations of Figures \ref{compP2} and \ref{compP3} were performed with \VD{a stopping criterion} $\epsilon=\|\bold{Au}-\bold{b}\|/\|\bold{b}\|$ equal to $10^{-3}$ \citep{Wilkinson_1963_REA}. Figure \ref{comptol} shows that this minimum tolerance is the optimal one as going beyond doesn't affect the precision of the numerical solution. This value of $\epsilon$ is also consistent with the analysis of \citet[][ Their Figure 3]{Sourbier_2011_GEOP}. Therefore, $\epsilon=10^{-3}$ will be used in the next section for the preconditioner assessment.
\begin{figure}[ht!]
\begin{center}
\includegraphics[width=8cm,clip=true]{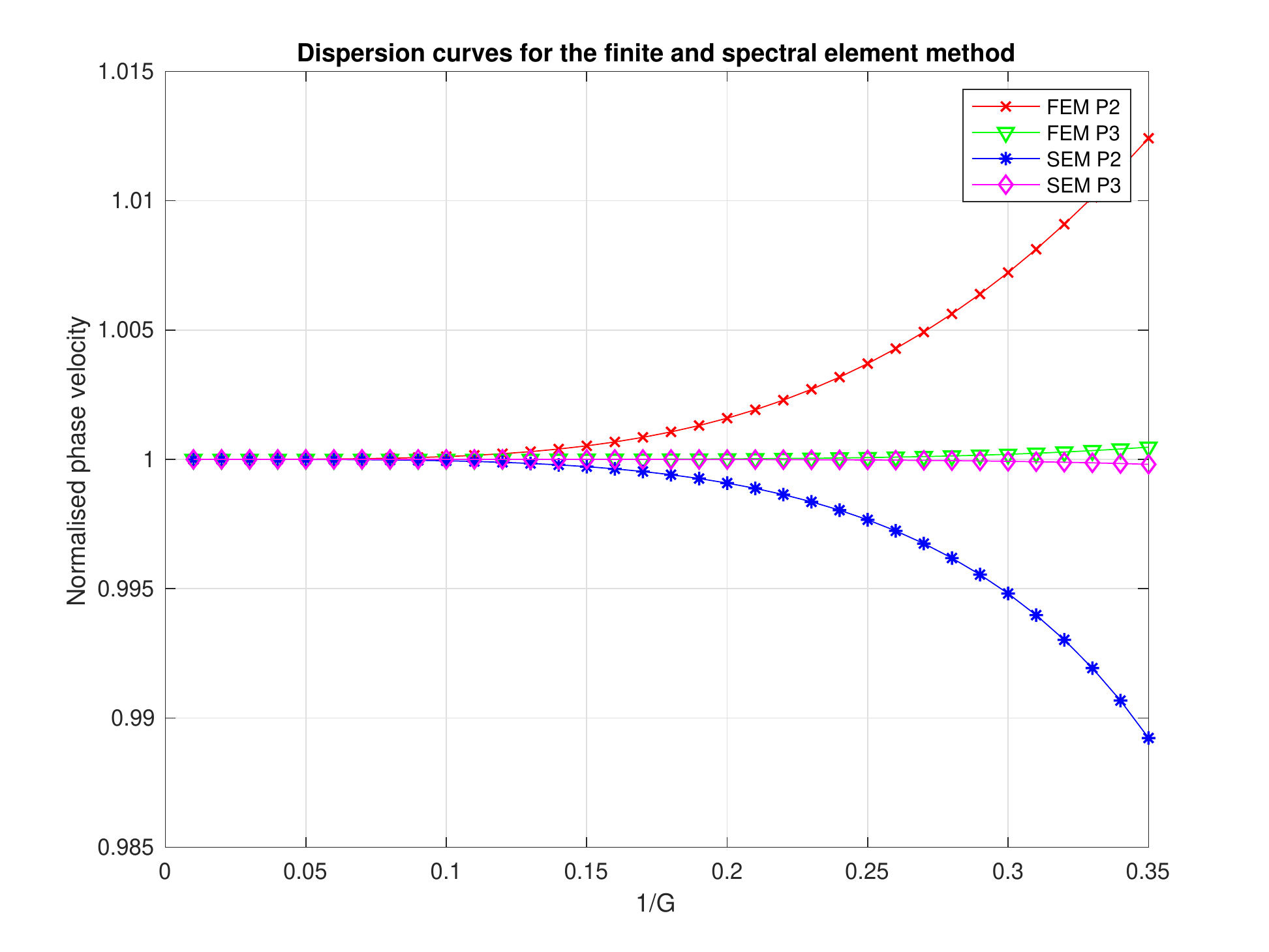}
\caption{Dispersion curves for P2 and P3 spectral and classical finite elements. $G$ denotes the number of grid points per wavelength. The normalized phase velocity is the ratio between the numerical phase velocity and the wavespeed.}
\label{disp}
\end{center}
\end{figure}
\begin{figure}[ht!]
\begin{center}
\includegraphics[width=7.5cm,clip=true]{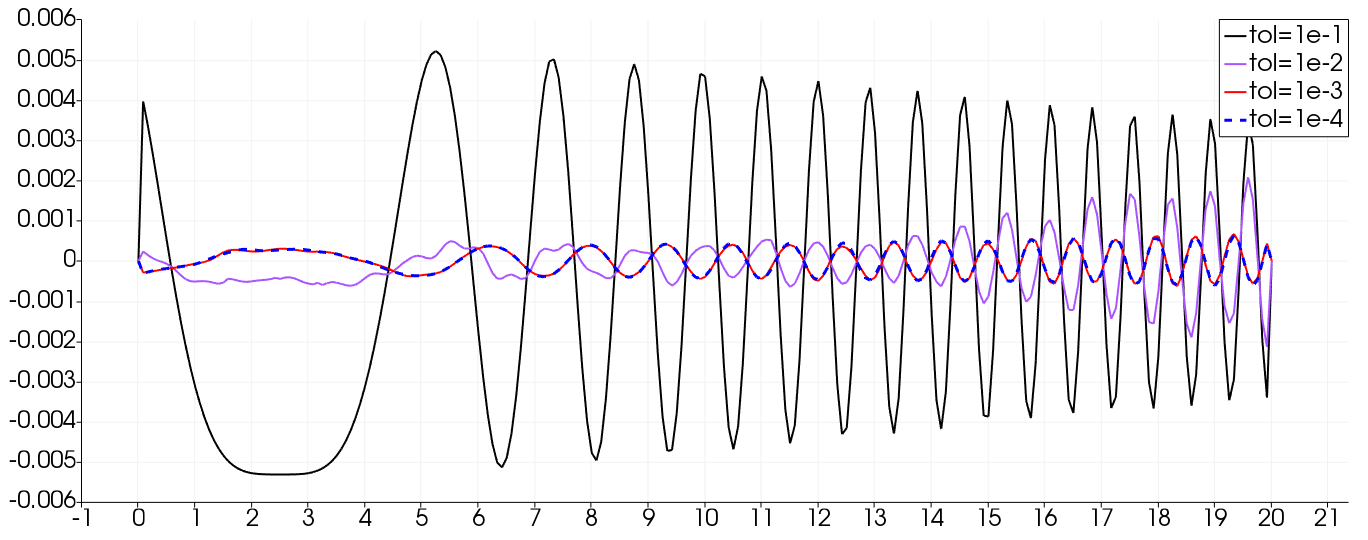}
\caption{Optimal backward error criterion. P3 FE solution with 5 ppwl and different tolerance ($\epsilon = 10^{-1}, 10^{-2}, 10^{-3}, 10^{-4}$). No significant improvement is shown for $\epsilon$ smaller than $10^{-3}$.}
\label{comptol}
\end{center}
\end{figure}
\vspace{-0.7cm}
\section{Domain decomposition preconditioner}
\vspace{-0.3cm}
We now review the preconditioner that we use to solve efficiently the linear system (\ref{eq1}).
A well-known iterative solver for this type of indefinite linear systems is the Krylov subspace Generalized Minimal RESidual Method (GMRES) \citep{Saad_2003_IMS}. However, the Helmholtz operator requires efficient preconditioning which can be done by domain decomposition \citep[][ section 2.2.1]{Dolean:2015:IDD}. \\
In this study, we solve system \eqref{eq1} with a two-level domain decomposition preconditioner $\bold{M}^{-1}$
\begin{equation}
\label{2lvl}
\bold{M}^{-1}  = \bold{M}^{-1}_1 (I - \bold{A} \bold{Q}) + \bold{Q}, \quad \text{with } \bold{Q} = \bold{Z} \bold{E}^{-1} \bold{Z}^T, \quad \bold{E} = \bold{Z}^T \bold{A} \bold{Z},  \\
\end{equation}
where $\bold{M}^{-1}_1$ is the one-level domain decomposition preconditioner called Optimized Restricted Additive Schwarz (ORAS) and $\bold{Z}^T$ is the interpolation matrix from the finite element space defined on $\Gamma$ onto a finite element space defined on a coarse mesh $\Gamma_H$. The construction of the domain decomposition preconditioner is described in detail in \citet{Bonazzoli:2019:ADD}. Let $\left\{\Gamma_i\right\}_{1 \le i \le N_d}$ be an overlapping decomposition of the mesh $\Gamma$ into $N_d$ subdomains. Let $\left\{\bold{A}_i\right\}_{1 \le i \le N_d}$ denote local Helmholtz operators with absorbing (or transmission) boundary conditions at the subdomain interfaces. The one-level ORAS preconditioner is
\begin{equation}
\bold{M}^{-1}_1 = \sum_{i=1}^{N_d} \bold{R}_i^T \bold{D}_i \bold{A}_i^{-1} \bold{R}_i,
\label{oras}
\end{equation}
where $\left\{\bold{R}_i\right\}_{1 \le i \le N_d}$ are the Boolean restriction matrices from the global to the local finite element spaces and $\left\{\bold{D}_i\right\}_{1 \le i \le N_d}$ are local diagonal matrices representing the partition of unity.\\
The key ingredient of the ORAS method is that the local matrices $\left\{\bold{A}_i\right\}_{1 \le i \le N_d}$ incorporate more efficient boundary conditions (i.e. absorbing boundary conditions) than in the standard RAS preconditioner based on local Dirichlet boundary value problems.\\
\VD{The coarse problem $\bold{E}$ in~\eqref{2lvl} is also solved iteratively by performing $10$ GMRES iterations with a one-level ORAS preconditioner}. We use the same spatial subdomain partitioning for the coarse and fine meshes. Each computing core is assigned to one spatial subdomain and holds the corresponding coarse and fine local matrices. Each application of the global preconditioner $\bold{M}^{-1}$ relies on local concurrent subdomain solves on the coarse and fine levels, which are performed by a direct solver. This hybrid direct/iterative solver requires careful strong scalability analysis to achieve the best compromise between parallel efficiency and memory storage.\\
%
\vspace{-0.7cm}
\section{Numerical results}
\begin{figure}[ht!]
\begin{center}
\includegraphics[width=7cm]{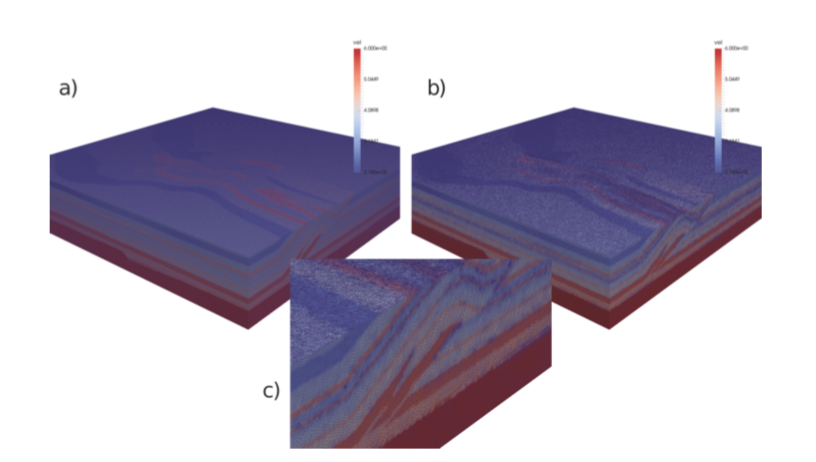}
\caption{Meshing of Overthrust model. (a) Regular and  (b) adaptive tetrahedral meshes.}
\label{over1}
\end{center}
\end{figure}

\begin{figure}[ht!]
\begin{center}
\includegraphics[width=8cm]{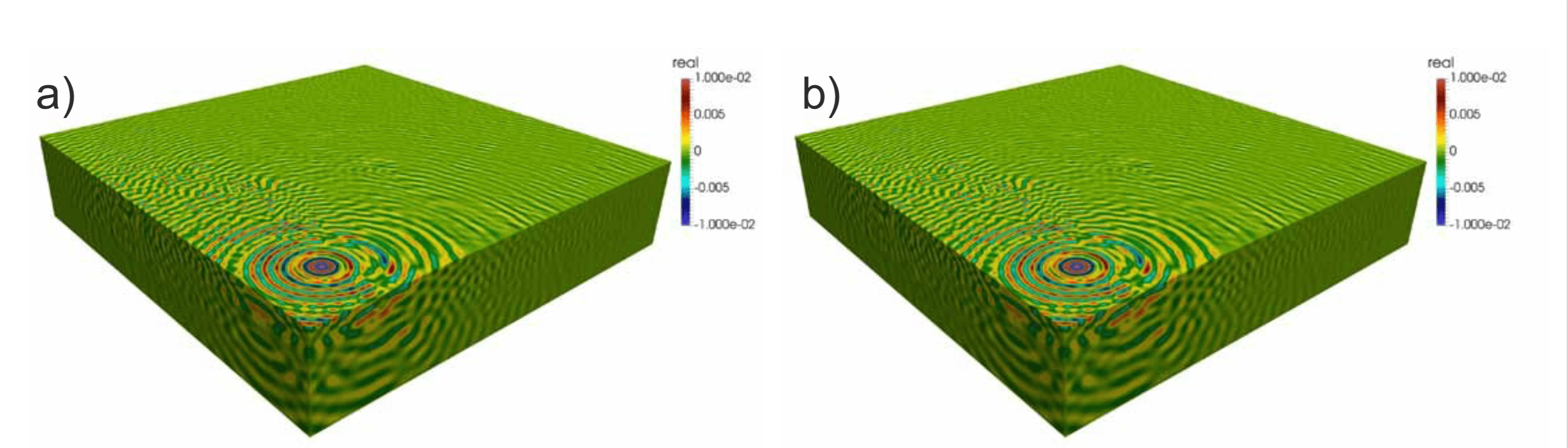}
\caption{10~Hz monochromatic wavefields computed in (a) regular and (b) adaptive tetrahedral meshes. (c) Zoom of (b).}
\label{over2}
\end{center}
\end{figure}

\vspace{-0.3cm}
\VD{The two-level solver} is implemented using the high-performance domain decomposition library HPDDM\footnote{\url{http://github.com/hpddm/hpddm}} (High-Performance unified framework for Domain Decomposition Methods)  \citep{Jolivet:2013:SDD}.
We assess the solver on the Ir\`ene supercomputer of TGCC\footnote{\url{http://www-hpc.cea.fr}} with the 3D $20 \times 20 \times 4.65$ km SEG/EAGE Overthrust model \citep{Aminzadeh_1997_DSO}. We perform wave simulation with P3 finite elements on regular and adaptive tetrahedral meshes (Fig. \ref{over1}) for the 5~Hz, 10~Hz and 20~Hz frequencies (Tab.~\ref{tab_over}) \VD{in double and single precision}. The average length of the element edges is set to 5 nodes per minimum wavelength on the regular tetrahedral mesh, and 5 nodes per local wavelengths in the adaptive tetrahedral mesh (2.5 for the coarser mesh used in the two-level method).
We use a homogeneous Dirichlet boundary condition at the surface and first-order absorbing boundary conditions along the other five faces of the model. The source is located at (2.5,2.5,0.58) km.  For weak scalability analysis, we keep $\#$dofs per subdomain roughly constant from one frequency to the next (Tab.~\ref{tab_over}). The $h$-adaptivity in the unstructured tetrahedral mesh decreases $\#$dofs relative to the regular mesh by a factor of 2.07. The stopping tolerance $\epsilon$ for GMRES is set to $10^{-3}$. The consistency between the 10~Hz wavefields computed in the regular and adaptive tetrahedral meshes is shown in Fig.~\ref{over2}. 

\begin{figure}[ht!]
\begin{center}
\includegraphics[width=5cm]{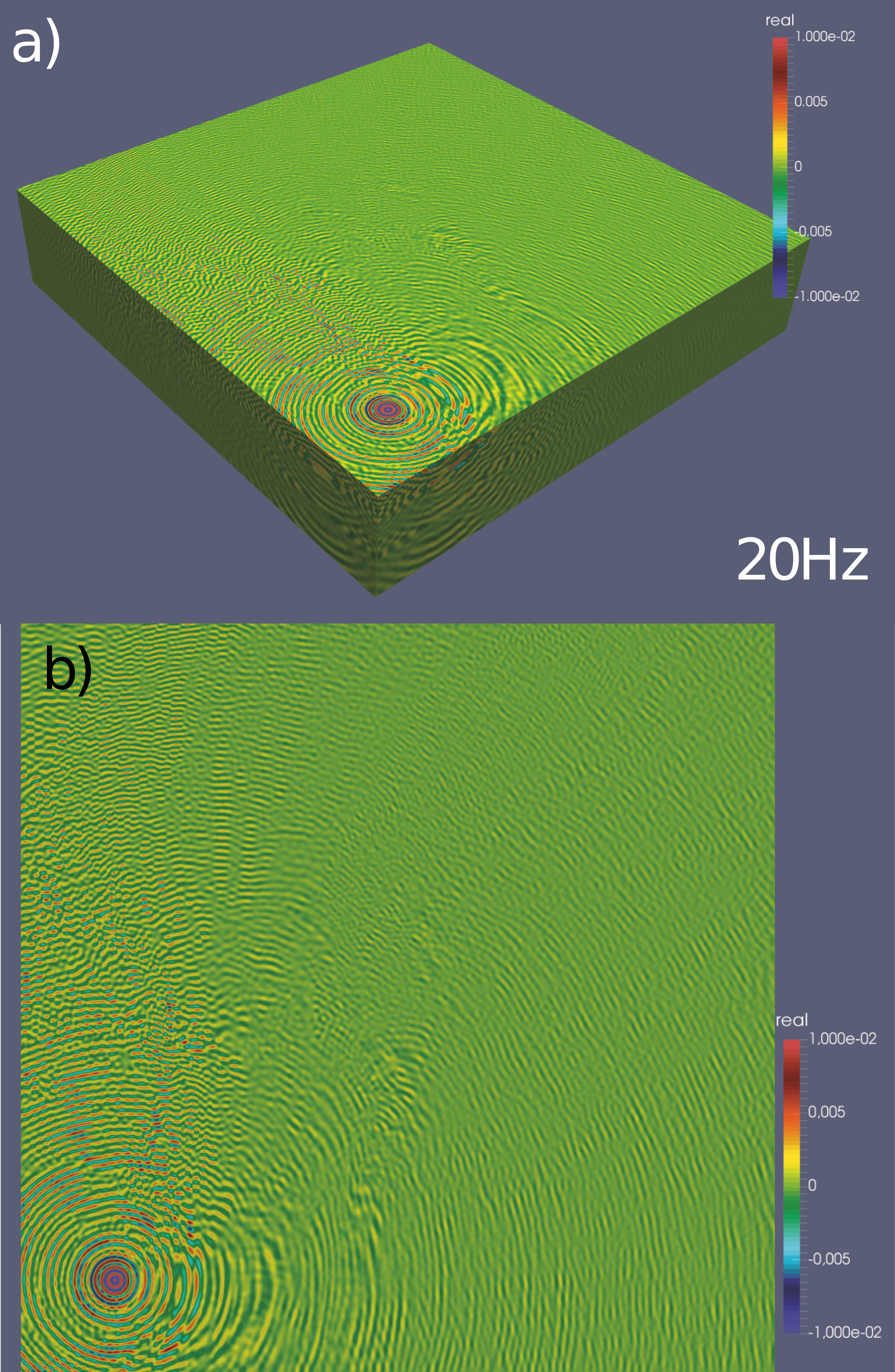}
\caption{20~Hz wavefield in the adaptive tetrahedral mesh.}
\label{over3}
\end{center}
\end{figure}

First, we carry out a set of numerical simulations at 5~Hz on the regular mesh in order to illustrate the benefits of performing computations in single precision arithmetic (versus double precision), as well as using an approximate factorization for the fine local matrices to apply $\bold{A}_i^{-1}$ in~\eqref{oras}. More precisely, we compare incomplete Cholesky factorization (ICC) to complete Cholesky factorization performed by Intel MKL PARDISO. The experiments are performed on 1060 cores with P3 finite elements and 5 ppwl, resulting in 74 million dofs. Results are reported in Tab.~\ref{tab_prec}. First, we can see that performing the whole computation in single precision instead of double precision yields a speedup of about 1.4 for the solution phase. The number of GMRES iterations is the same, there is no loss of accuracy or additional numerical instability. Additionally, the setup phase is drastically reduced (speedup 1.8) when performing Cholesky factorization in single precision. Second, we can see that using an incomplete Cholesky factorization for the fine local matrices yields a speedup of about 1.6 with respect to complete factorization, once again with no effect on the number of GMRES iterations. Moreover, the memory savings are pretty significant: with complete Cholesky factorization we run out of memory with 768 cores, while the simulation runs on 265 cores using ICC.
In the rest of this paper, the experiments are performed in single precision and using incomplete Cholesky factorization for the fine local matrices. Timings for the adaptive tetrahedral mesh are around two times smaller than those obtained on the regular mesh (Tab.~\ref{tab_over}). The simulation at 20~Hz on the adaptive mesh (Fig.~\ref{over3}) involves 2,285 millions of dofs and requires 16,960 cores. The elapsed time achieved by the 2-level preconditioner is 15s and 37s for 10~Hz and 20~Hz respectively (Tab.~\ref{tab_over}). 

%
%
\vspace{-0.5cm}
\section{Conclusion}
\vspace{-0.3cm}
We have proposed an efficient and accurate forward engine for 3D frequency-domain FWI from ultra-long offset stationary-recording survey. Two key ingredients are the $h$-adaptive tetrahedral P3 finite-element discretization to optimize the number of d.o.f in the domain and conform the mesh to complex boundaries, and a massively-parallel preconditioned iterative solver for efficient solution of Helmholtz problem. \VD{For a comparison with the one-level method and the management of multiple RHS in the case of P2 discretization, see \citet{Dolean:2020:IFS}}. Multi-RHS processing can be further improved with the Krylov subspace recycling method GCRO-DR \citep{Parks_2006_RKS} and its block variant, which are already implemented in the HPDDM library and have been applied successfully for medical imaging based on a multi-antenna microwave device \citep{Tournier:2019:MTI}. Extension to visco-elastic media is part of a future work.
%

%
%
\vspace{-0.1cm}
\textbf{Acknowledgments:}\\
This study was granted access to the HPC resources of SIGAMM\footnote{\url{http://crimson.oca.eu}} and CINES/IDRIS under the allocation 0596 made by GENCI. This study was partially funded by the WIND consortium\footnote{\url{https://www.geoazur.fr/WIND} sponsored by Chevron, Shell and Total}.

\begin{table}[ht!]
\begin{center}
\begin{tabular}{|c|c|c|c|c|}
\hline
\multicolumn{5}{|c|}{\bf{Cartesian grid, \; f = 5Hz}} \\ \hline
precision & fine local solver & $\#$it & setup & GMRES \\ \hline
double & Cholesky & 10 & 92.5s & 15.5s \\ \hline
double & ICC & 10 & 30.2s & 8.9s \\ \hline
single & Cholesky & 10 & 50.3s & 10.3s \\ \hline
single & ICC & 10 & 25.8s & 6.3s \\ \hline
\end{tabular}
\end{center}
\caption{Comparison between Cholesky and incomplete Cholesky factorization (ICC) of local matrices at the fine level, and single versus double precision arithmetic for the whole computation, at 5Hz with P3 finite elements and 5 ppwl (74M dofs) on 1060 cores. $\#$it: number of iterations. Elapsed time in seconds for the setup phase (assembly and factorization of local matrices) and solution phase with GMRES ($\epsilon=10^{-3}$).}
\label{tab_prec}
\end{table}

\begin{table}[ht!]
\begin{center}
\begin{tabular}{|c|c|c|c|c|c|}
\hline
\multicolumn{6}{|c|}{\bf{Regular mesh}} \\ \hline
Freq (Hz) & $\#$core & $\#$elts  (M)& $\#$dofs (M)   & $\#$it & GMRES \\ \hline
{\bf{5}} & 265 & 16 & 74 & 7 & 16s \\ \hline
{\bf{10}} & 2,120 & 131 & 575 & 15 & 33s \\ \hline
\multicolumn{6}{|c|}{\bf{Adaptive mesh}} \\ \hline
Freq (Hz) & $\#$core & $\#$elts  (M)& $\#$dofs (M)   & $\#$it & GMRES \\ \hline
{\bf{10}} & 2,120 & 63 & 286 & 14 & 15s \\ \hline
{\bf{20}} & 16,960 & 506 & 2,285 & 30 & 37s \\ \hline
\end{tabular}
\end{center}
\caption{Statistics of simulation in regular and adaptive tetrahedral meshes. $Freq (Hz)$: frequency; $\#$core: number of cores; $\#$elts: number of elements; $\#$dofs: number of degrees of freedom; $\#$it: iteration count. Elapsed time (seconds) in GMRES.}
\label{tab_over}
\end{table}

\onecolumn
\bibliographystyle{seg}  
\newcommand{\SortNoop}[1]{}

\end{document}